\documentclass[aps,prd,showpacs,eqsecnum,twocolumn,superscriptaddress]{revtex4}

\usepackage{latexsym} \usepackage{amssymb} \usepackage{amsfonts}
\usepackage{amsmath} \usepackage{bm} \usepackage[dvips]{graphicx}
\usepackage{color}
\usepackage{subfigure} \usepackage{times} \usepackage{units}
\usepackage{hyperref}
\usepackage{bm}
\usepackage[utf8x]{inputenc} \usepackage{amssymb,amsmath}
\usepackage{graphicx}

\begin{document}
\title{Total-variation methods for gravitational-wave denoising: performance tests on Advanced LIGO data}

\author{Alejandro Torres-Forn\'e}
\affiliation{Departamento de Astronom\'{\i}a y Astrof\'{\i}sica, Universitat de Val\`encia,
Dr. Moliner 50, 46100, Burjassot (Val\`encia), Spain} 
 
\author{Elena \surname{Cuoco}}
\affiliation{European Gravitational Observatory (EGO), Via E Amaldi, I-56021 Cascina, Italy}
\affiliation{Scuola Normale Superiore (SNS), Piazza dei Cavalieri, 7, 56126 Pisa PI, Italy}
\affiliation{ Istituto Nazionale di Fisica Nucleare (INFN) Sez. Pisa Edificio C - Largo B. Pontecorvo 3, 56127 Pisa, Italy.}

\author{Antonio \surname{Marquina}}\affiliation{Departamento de
 Matem\'atica Aplicada, Universitat de Val\`encia,
 Dr. Moliner 50, 46100, Burjassot (Val\`encia), Spain} 
 
\author{Jos\'e A. \surname{Font}}\affiliation{Departamento de
 Astronom\'{\i}a y Astrof\'{\i}sica, Universitat de Val\`encia,
 Dr. Moliner 50, 46100, Burjassot (Val\`encia), Spain}
 \affiliation{Observatori Astron\`omic, Universitat de Val\`encia, C/ Catedr\'atico 
 Jos\'e Beltr\'an 2, 46980, Paterna (Val\`encia), Spain} 
 
\author{Jos\'e M. \surname{Ib\'a\~nez}}\affiliation{Departamento de
 Astronom\'{\i}a y Astrof\'{\i}sica, Universitat de Val\`encia,
 Dr. Moliner 50, 46100, Burjassot (Val\`encia), Spain} 
 \affiliation{Observatori Astron\`omic, Universitat de Val\`encia, C/ Catedr\'atico
 Jos\'e Beltr\'an 2, 46980, Paterna (Val\`encia), Spain}


\begin{abstract}
We assess total-variation methods to denoise gravitational-wave signals in real noise conditions, by injecting numerical-relativity waveforms from core-collapse supernovae and binary black hole mergers in data from the first observing run of Advanced LIGO. This work is an extension of our previous investigation where only  Gaussian noise was used. Since the quality of the results depends on the regularization parameter of the model, we perform an heuristic search for the value that produces the best results. We discuss various approaches for the selection of this parameter, either based on the optimal, mean, or multiple values, and compare the results of the denoising upon these choices. Moreover, we also present a machine-learning-informed approach to obtain the Lagrange multiplier of the method through an automatic search. Our results provide further evidence that total-variation methods can be useful in the field of Gravitational-Wave Astronomy as a tool to remove noise.
\end{abstract}
\maketitle

\section{Introduction}
\label{section:intro}

The observation of gravitational waves from coalescing binary black holes (BBH) during the first Advanced 
LIGO~\cite{AdvLIGO} observing run (O1) marked the commencement of gravitational-wave astronomy~\cite{GW150914-prl,GW151226-prl}. After a period of commissioning, the two LIGO detectors started the second observing run (O2) by the end of 2016, with the European detector Advanced Virgo \cite{AdvVirgo} joining on August 2017. O2 was an overwhelming success. In addition to the observation of three new BBH mergers~\cite{GW170104,GW170608,GW170814}, the latter simultaneously observed by the three detector network, it also accomplished the first observation of gravitational waves from a binary neutron star (BNS) 
merger~\cite{GW170817}. Unlike BBH events, BNS mergers emit electromagnetic signals across the entire spectrum. Those were detected by dozens of telescopes, opening the field of multi-messenger astronomy~\cite{MMA}. 

During O2, the BNS observational range of Advanced LIGO was as large as $\sim$100 Mpc. However, due to their intrinsic weakness, signals from most astrophysical sources within such a large volume are likely to remain in the limit of detectability. A careful analysis of the collected gravitational-wave data is therefore essential to ensure progress in spite of the conspicuous instrumental noise of the detectors. Actual gravitational-wave signals may be misinterpreted as artificial noise transients (glitches), which requires their precise identification and eventual veto. 

Noise removal is a long-standing, major effort in gravitational-wave data analysis, and specific algorithms have been developed for every type of signal. For coalescing compact binary (CBC) signals, such as the existing sample of observations, the inspiral signal can be observed by either targeting a broad range of generic transient signals or by correlating the data with analytic waveform templates from general relativity and maximizing such correlation with respect to the waveform parameters~\cite{Schutz:2009,Usman:2016}. Identification is challenged for events with low signal-to-noise ratio (SNR) due to the non-stationarity and non-Gaussanity of the detector noise. Matched-filtering is impractical for well-modeled but continuous sources, like spinning (isolated) neutron stars, due to the large computational resources it would require. Cross-correlation and coherent detection methods~\cite{bose:2000,Palomba:2012} are the choice for these sources. In addition to CBC and continuos sources, there also exist other sources that produce gravitational-wave transients (or bursts), with core-collapse supernova (CCSN) signals being a paradigmatic example. Those can only be modeled imperfectly, as the computational requirements for obtaining their waveforms from numerical relativity simulations are significant and the intrinsic parameter space is larger than for CBC signals. Recently, coherent approaches over a network of detectors have proven very effective~\cite{Thrane:2015,Klimenko:2016}, increasing the detection confidence of long-duration burst signals (above several seconds). In contrast, short-duration bursts are more affected by detector glitches and specific pipelines have been developed to differentiate between signals and noise transients, namely BayesWave~\cite{Littenberg:2016}, either standalone or in combination with coherentWaveBurst~\cite{kanner:2016}, and oLIB~\cite{Lynch:2016}. Approaches to estimate physical parameters and to reconstruct burst signal waveforms from noisy environments have also been put forward by~\cite{Rover:2009,Engels:2014,powell:2016,powell:2017}. 

Methods based on machine learning (ML) offer a promising alternative to current approaches, having already shown optimal performance for many tasks, like classification and regression, and in many scientific disciplines~\cite{Goodfellow:2016}. ML techniques have been recently applied for gravitational-wave astronomy as well~\cite{Torres:2016,Razzano:2018,Gabbard:2018,George:2018}. In a previous paper~\cite{Torres:2014} we assessed total-variation (TV) algorithms for denoising gravitational-wave signals. TV methods are based on $L_1$ norm minimization and have been mainly employed in the context of image processing, where they constitute the best approach to solve the so-called Rudin-Osher-Fatemi denoising model~\cite{Rudin:1992}. Our first investigation~\cite{Torres:2014} was limited to denoise gravitational waves embedded in additive {\it Gaussian} noise. We showed that noise can be successfully removed with TV techniques, irrespective of the signal morphology or astrophysical origin. In the current paper we take a further step in the assessment of TV-methods for gravitational-wave astronomy, using actual noise from the detectors instead of the idealized non-white Gaussian noise employed in our previous study. To this aim, we inject numerically generated signals (from BBH mergers and CCSN) into the data collected by the Advanced LIGO detector during the O1 observing run that extended from 2015 Sep 12th to 2016 Jan 19. Our goal is to test whether TV-methods can effectively reduce noise in the conditions found with real data.

The paper is organized as follows: In Section~\ref{section:TV_methods} we summarize the mathematical framework on which TV methods are based. Section~\ref{section:setup} briefly explains the whitening method we employ to remove noise lines and other artefacts, and the waveform catalogs we use to test our algorithm. In Section \ref{section:estimation} we discuss the determination of the regularization parameter which produces the best results for the sources considered. The main results of our study are presented in Section~\ref{section:results}. Finally, a summary is provided in Section~\ref{section:summary}

\section{Overview of the method} 
\label{section:TV_methods}

TV methods are based on the concept of TV-norm regularization, 
introduced by~\cite{Rudin:1992} in 1992 as a procedure to
clean noisy signals. Starting from the classical linear degradation model, 
$f=u+n$, where a noisy signal $f$ is built from a signal 
$u$ and some additive noise $n$, we assume as white Gaussian noise
with zero mean and variance $\sigma^2$.

The variational approach to recover $u$ (clean signal) from $f$ 
(the observed signal) and some information about the noise is
to priorize signals $u$, through the minimization of $R(u)$ a convex energy,
called regularizer, subject to the constraint that the square of the 
$||f-u||_{\text{L}_2}^2$ matches the variance of the noise, $\sigma^2$.
Applying Tihonov theorem the constrained variational problem can be 
written in general as an unconstrained 
minimization problem, by introducing a positive Lagrange multiplier $\mu>0$:
\begin{equation}
 \label{eq:unconstrainL2}
 u=\underset{u} {\text{argmin}}\left\{R(u)+\frac{\mu}{2} \, ||f-u||_{\text{L}_2}^2\right\}~.
 \end{equation}

This energy has two main terms: $R(u)$ is called the {\it regularization term}
that rules out the signals with large values of $R(u)$. 
The second term $||f-u||_{\text{L}_2}^2$ is called the {\it fidelity term} 
and controls the degree of similarity between the solution $u$ and 
the noisy signal $f$ by computing the square of the $\text{L}_2$-norm. 
Both terms are weighted by a Lagrange multiplier $\mu >0$ so that, 
when it has a small value, the relative weight of the fidelity term is small 
and the solution is highly regularized. In contrast, 
when the value of $\mu$ is high, the solution is dominated 
by the fidelity term and $u$ is similar to $f$.

In this paper we shall use as regularizing energy $R(u)$ either the  
$\text{L}_1$ of the signal (LASSO) (see Tibshirani~\cite{lasso}) or the $\text{L}_1$ of
its gradient (see ~\cite{Rudin:1992}), which favors {\it sparse} solutions, 
i.e.~very few nonzero components of the solution or its gradient. 
In addition the algorithm to find $\text{L}_1$-norm minimizers 
is extremely efficient despite this norm is not differentiable.

%
%
%

\begin{figure*}
 \centering

 {\includegraphics[width=0.49\textwidth]{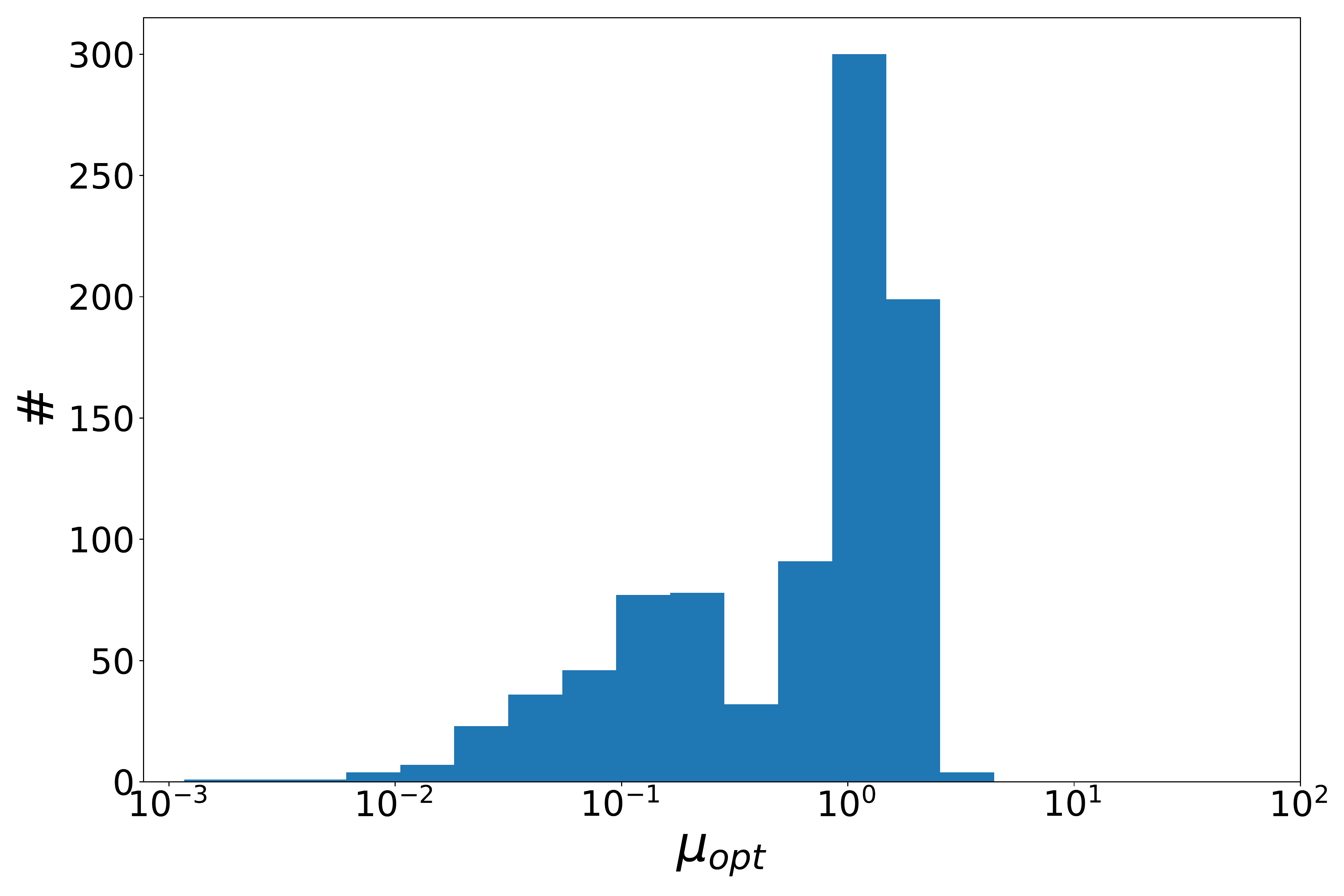}}
 {\includegraphics[width=0.49\textwidth]{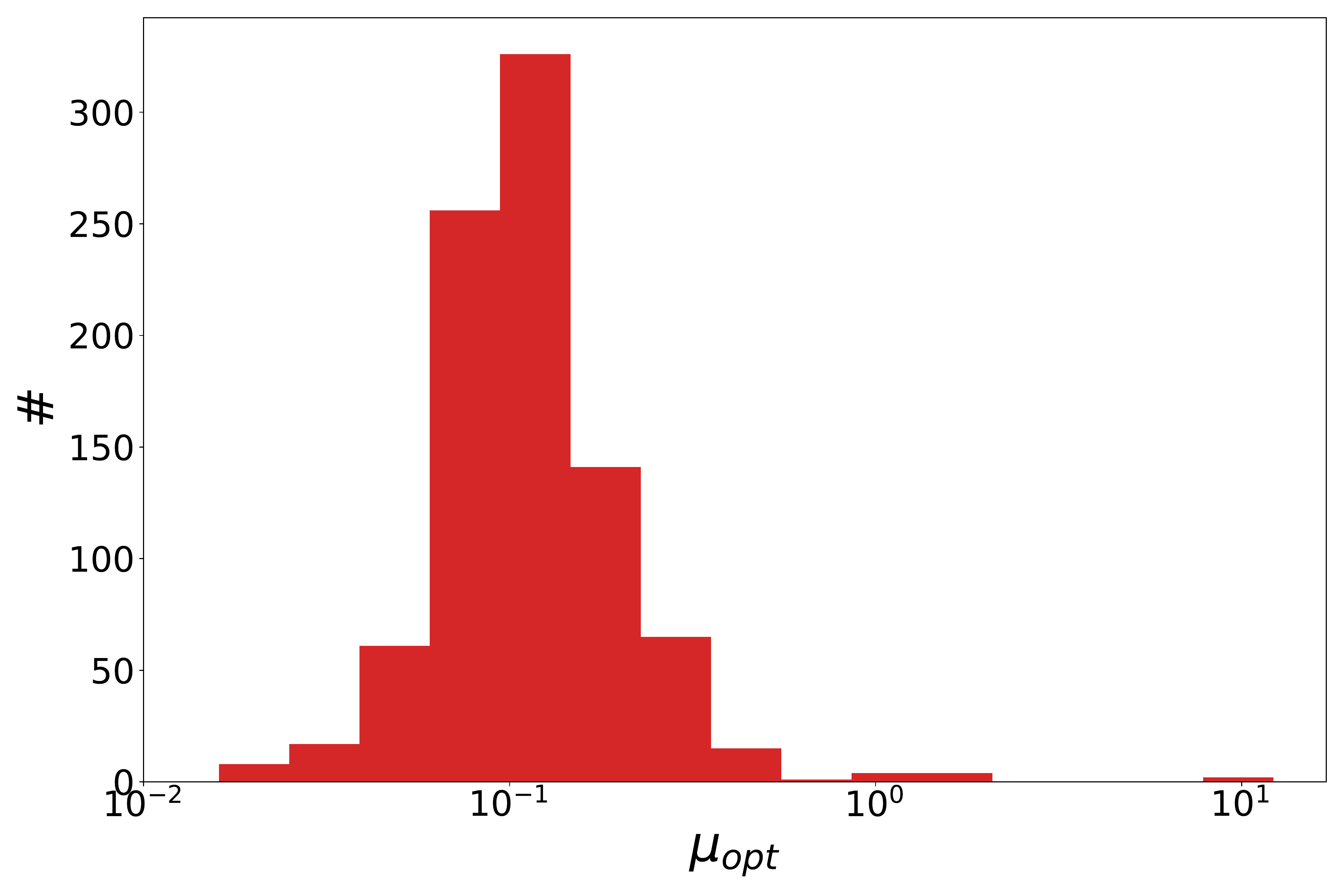}}

 \caption{Histograms of the optimal values of $\mu$ for the training signals from the core-collapse catalog (left panel) and BBH catalog (right panel) at three different distances in ten different gps times. }
  \label{fig:histograms}
\end{figure*}

Rudin, Osher and Fatemi in his pioneering paper \cite{Rudin:1992} 
proposed the use of the $\text{L}_1$ norm of the gradient for the regularizing
energy. This specific formulation of the variational 
problem~(\ref{eq:unconstrainL2}) is called ROF model 
and reads
\begin{equation}
\label{eq:rof}
u=\underset{u} {\text{argmin}}\left\{\mathrm{TV}(u)+\frac{\mu}{2}||u-f||_{\text{L}_2}^2\right\}~.
\end{equation}
Since the ROF model uses the TV-norm the solution is the only one with the {\it sparsest} gradient. 
Thus, the ROF model reduces noise by {\it sparsifying} the gradient of the signal and 
avoiding spurious oscillations (ringing) on the solution signal.

The associated Euler-Lagrange equation of the ROF model is given by
\begin{equation}
\label{eq:EL_TV}
\nabla \cdot \frac{\nabla u }{|\nabla u|}+ \mu (f-u) = 0~.
\end{equation}
This equation becomes singular when $|\nabla u| = 0$. To avoid this, 
the following regularized TV-norm was used in~\cite{Torres:2014} and 
we will call it the regularized ROF (rROF) algorithm, which allows to 
obtain an approximate solution of the ROF model by smoothing the total variation energy. 
The TV functional is slightly perturbed as
\begin{equation}
\mathrm{TV}_{\beta}(u):= \int \sqrt{|\nabla u|^2+\beta}
\end{equation}
where $\beta$ is a small positive parameter. Therefore, the regularized ROF model reads
\begin{equation}
\label{eq:rof_beta}
u=\underset{u} {\text{argmin}}\left\{\mathrm{TV}_{\beta}(u)+\frac{\mu}{2}||u-f||_{\text{L}_2}^2\right\}~.
\end{equation}
Assuming homogeneous Neumann boundary conditions, Eq.~(\ref{eq:rof_beta}) 
becomes a nondegenerate second order nonlinear elliptic differential equation 
whose solution is smooth, (for details see section II in~\cite{Torres:2014}).

In this paper we use a regularized ROF algorithm by solving the 
associated Euler-Lagrange equation of an energy that includes a 
smoothed TV norm (see~\cite{Torres:2014} for details). 
iThe novelty here consists of using the rROF algorithm
as a building block of an iterative procedure, called
Bregman iterative procedure (see~\cite{Osher:2005}). 
runs the scale space from the the solution of the 
regularizer TV-model, using a very small
value of the Lagrange multiplier, to the processed signal. 
Roughly speaking, we first choose the regularization parameter 
$\mu$ equal to a constant value $\mu_0$, which is smaller than 
the optimal value needed to obtain a denoised signal 
by direct application of the rROF algorithm. 
The value of $\mu_0$ is kept fixed through all the iterations. 
Next, we compute $u_1$ by solving
\begin{eqnarray}
u_1&=&\underset{u} {\text{argmin}}\left\{\mathrm{TV}_{\beta}(u)+\frac{\mu_0}{2}||u-f||_{\text{L}_2}^2\right\}~,
\\
f&=&u_1+v_1~,
\end{eqnarray}
where $v_1$ is the residual.
Then, we apply again the rROF algorithm using the same $\mu_0$ and 
taking as input signal $f+v_1$ to obtain $u_2$. We thus have
\begin{eqnarray}
f+v_1&=&u_2+v_2~.
\end{eqnarray}
Applying this procedure for an arbitrary number of times $n$ we obtain 
a sequence of signals $u_n$ for $n=1, \cdots$ such that
\begin{eqnarray}
f+v_{n-1}&=&u_n+v_n~.
\end{eqnarray}
The iteration stops when some discrepancy principle is satisfied, 
i.e.~when the square of the $\text{L}_2$-norm 
of the residual matches the variance of the noise.
In practice, however, the variance of the noise is not available 
and we have to resort to some other termination criterion.
We refer to our original paper~\cite{Torres:2014} for details.

Our test over the signals examples show that a tolerance of $10^{-3}$ 
for both the rROF algorithm and $10^{-2}$ for 
the iterative step are a good compromise between the accuracy on the results and computation speed.
The number of Bregman iterations are set to be at most a couple of iterations for the same reasons. 
These algorithm parameters will remain the same for the cases considered in this paper.

\section{Algorithm pipeline and data conditioning}
\label{section:setup}

Our previous work~\cite{Torres:2014} has shown that the rROF algorithm leads to satisfactory results for signals embedded in Gaussian noise. However, the noise of gravitational-wave detectors is non-Gaussian and non-stationary. For example, there are well-known, modeled sources of narrow-band noise, such as the electric power (at 60 Hz and higher harmonics), mirror suspension resonances or calibration lines (see Fig.~3 of~\cite{GW150914-prl}). For this reason, data must first be pre-conditioned to make the noise flat in frequency (a process known as whitening). To do so we make as few assumptions as possible. 
In this work we employ 10 chunks of data of 5 s each from the Advanced LIGO Livingston detector to inject the different signals. The GPS times are selected randomly over the entire O1 period. The sampling frequency is 16384 Hz.

We preprocess the data using the whitening procedure developed by~\cite{Cuoco:2001A,Cuoco:2001B}. This procedure uses an autoregressive (AR) model to transform the colored noise into white noise (see~\cite{Cuoco:2001A} for details). First we obtain the 3000 coefficients of the AR filter required by the whitening using 300 s of data at the beginning of the corresponding science segment of every signal injection. The whitening is applied in the time domain to every block of 5 s of data we use, in order to avoid the border problems associated with the transformations in the frequency domain. 

As in~\cite{Torres:2014} we apply the TV-method to two different types of gravitational-wave signals. The first type of waveforms are bursts from CCSN. We employ the waveform catalog of Dimmelmeier et al.~\cite{Dimmelmeier:2008}, who built a catalog of 128 waveforms from general relativistic simulations of rotating stellar core collapse to neutron stars. The simulations considered progenitors with high rotation rate and two tabulated, microphysical  equations of state (EoS). The second type of waveforms is based on the 174 numerical simulations from the inspiral and merger of BBH of Mrou\'e et al.~\cite{Mroue:2013} of which 167 cover more than 12 orbits and 91 represent precessing binaries. 

The rROF algorithm is coded in Fortran combined with a Python interface for plotting purposes. The algorithm is very efficient: the average time of 1000 runs of 3 s of data takes $\sim 16$ ms, computed in a single processor 3.5 GHz Intel Core i7 with 16 Gb of RAM. The iterative procedure, including the Bregman iteration, takes on average 0.5 s to perform the denoising of 3 s of data. 

\section{Estimation of the regularization parameter}
\label{section:estimation}

As already discussed in~\cite{Torres:2014} the denoising results strongly depend on the value of the regularization parameter $\mu$. If this value is too large the fidelity term in Eq.~(\ref{eq:rof}) dominates and the denoised signal is comparable to the original noisy signal $y$. On the contrary, if the value of $\mu$ is too small, it is the regularization term in Eq.~(\ref{eq:rof}) the dominant one and the amplitude of the resulting signal tends to zero. The existence of an {\it optimal} value of $\mu$ can be proven theoretically. However, this unique value is not equally appropriate for all possible cases one may consider (involving differences in the noise and/or in the signals) and must therefore be set empirically in practice. 
 
In this section we determine the interval of values of $\mu$ where satisfactory results are expected. This is similar to the analysis we performed in~\cite{Torres:2014} apart from the fact that we now use a logarithmic scale because the minimizer function converges faster than with a linear scale. For this reason, the regularization parameter used in the rROF algorithm is  $10^{\mu}$. The optimal value, $\mu_{\rm opt}$, is the one that gives the best results according to some suitable metric function applied to the denoised signal and to the original one. This function is used to measure the quality of the recovered signal. In~\cite{Torres:2014} we chose the peak signal-to-noise ratio (PSNR) as our quality estimator. In the present paper we assess the results of the iterative rROF algorithm using the structural similarity (SSIM) index, motivated by the quality assessment based on error sensitivity reported by~\cite{Wang:2004}. This estimator deviates from the traditional measures of error, which are based on the calculation of the absolute error, because it takes into account the structural information of both the original and the reconstructed signals. The SSIM index varies between 0 (minimum similarity) and 1 (maximum similarity) and is defined as,
\begin{equation}
\rm{SSIM}(x,y) = \frac{(2\mu_x\mu_y + c_1)(2\sigma_{xy} + c_2)}{(\mu_x^2 + \mu_y^2 + c_1)(\sigma_x^2 + \sigma_y^2 + c_2)},
\label{eq:ssim}
\end{equation}
where $c_1$ and $c_2$ are constants, $\mu_x$ ($\mu_y$) is the average of $x$ ($y$), $\sigma_x^2$ ($\sigma_y^2$) the variance of $x$ ($y$) and $ \sigma_{xy}$ the covariance of $x$ and $y$. 
The error provided by the SSIM index is used to determine the optimal value of the regularization parameter $\mu$ in each case.

We search for the optimal value of $\mu$, i.e. the one that maximizes the SSIM, injecting numerical relativity signals from the CCSN and BBH catalogs into O1 data. For the former we employ 30 different CCSN signals at three different distances, namely 5, 10 and 20 kpc. These distances are the same as were used in~\cite{powell:2016} and represent a reasonable example of distance for the signals in the Dimmelmeier catalog~\cite{Dimmelmeier:2008}. The injections are performed at 10 different random GPS times over the Advanced LIGO O1 data. With all this data, we obtain the histogram of optimal values of $\mu$ shown in the left panel of Fig.~\ref{fig:histograms}. We follow the same procedure for the BBH signals of~\cite{Mroue:2013}, but in this case the chosen distances are 400, 800 and 1000 Mpc, which are similar to the distances of the detected BBH events~\cite{GW150914-prl,GW151226-prl,GW170104,GW170608,GW170814}. The corresponding histogram is shown in the right panel of Fig.~\ref{fig:histograms}.

The mean value of $\mu$ for CCSN signals is $\bar{\mu}_{\rm opt}=-0.28$ with a standard deviation of $\sigma_{\rm opt} = 0.58$. In the case of BBH signals, the values are $\bar{\mu}_{\rm opt} = -0.95$ and $\sigma_{\rm opt} = 0.27$, respectively. The mean values of $\mu_{\rm opt}$ are different for both types of signals. This is expected since the two signals are very different and the conditions that apply for one type do not apply for the other. Specifically, for the BBH signals we have centred our analysis in the denoising of the very last cycles of the inspiral, the merger and the ring-down parts. This selection produces unsatisfactory denoising of the inspiral part of the signal as we show later.

Although the mean value of $\mu_{\rm opt}$ is different for both catalogs, Fig.~\ref{fig:histograms} also shows that partial overlap exists between the two distributions. This is expected since if we knew the variance of the noise it would be possible to determine the most appropriate value of the Lagrange multiplier for the fidelity term that corresponded to that variance. On the other hand, in a realistic situation the template is unknown which renders impossible to obtain $\mu_{\rm opt}$. Therefore, other strategies are required to determine the values of the regularization parameter that produce good results. For this reason, in this work we also try out and compare two different approaches based on the information provided by the histograms of Fig.~\ref{fig:histograms}. The first one is based on the mean value of $\mu$ for all waveforms. The second approach is to use the average of 20 different values sampled from a Gaussian distribution with the same mean and variance as the corresponding histograms.

\section{Results}
\label{section:results}

\subsection{Core-collapse supernova signals}

\begin{figure}
 \centering
 {\includegraphics[width=0.45\textwidth]{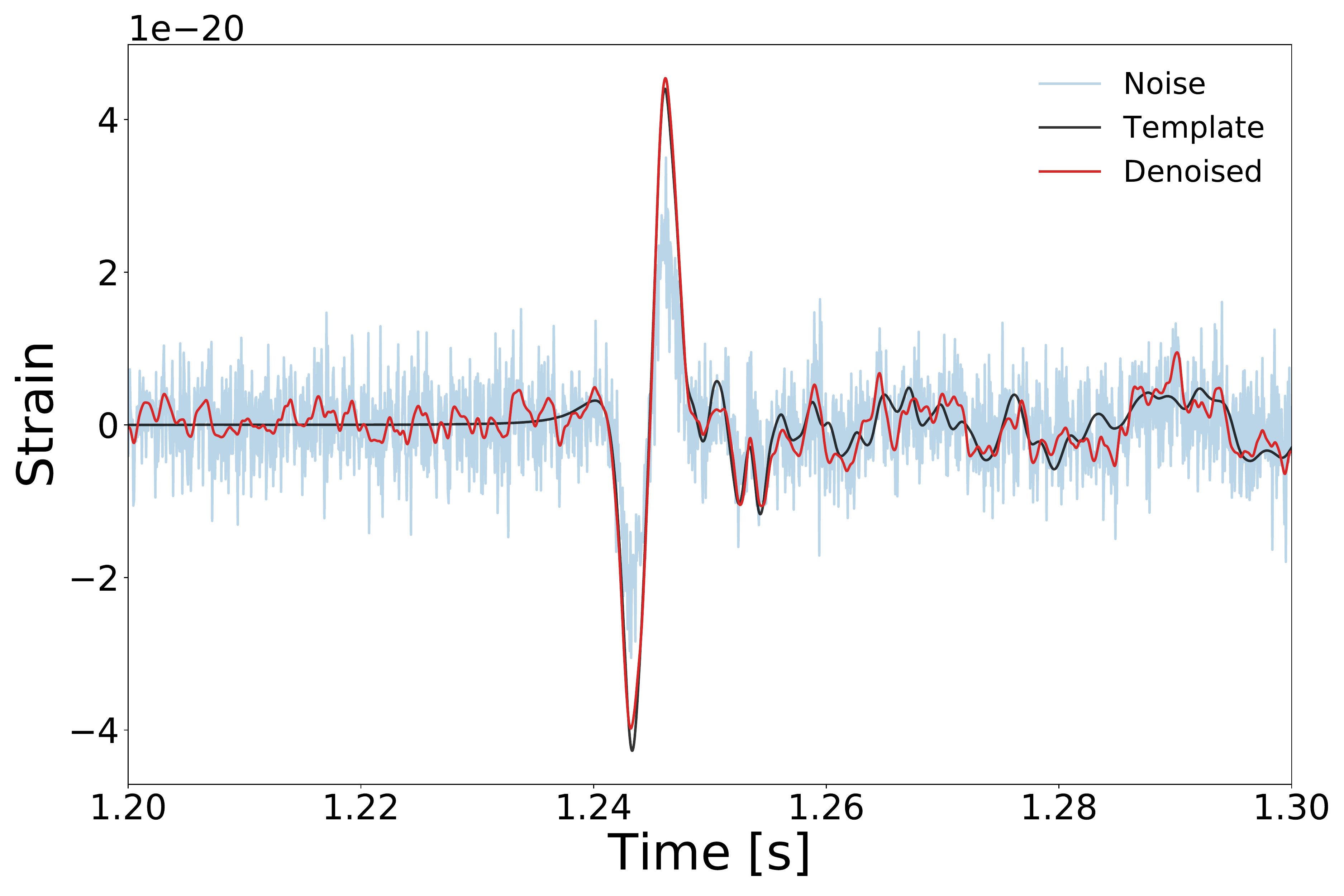}}
 {\includegraphics[width=0.45\textwidth]{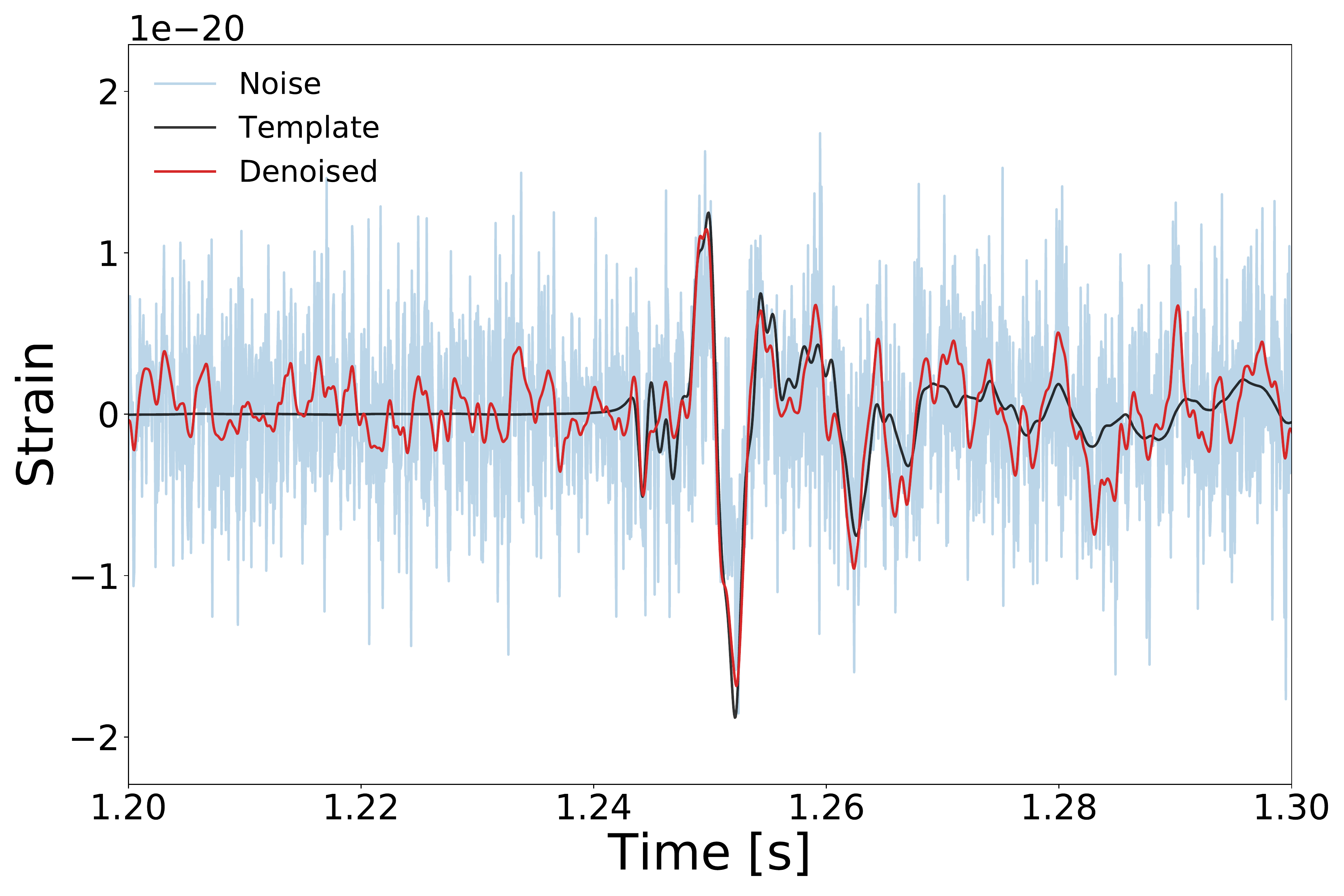}}
 {\includegraphics[width=0.45\textwidth]{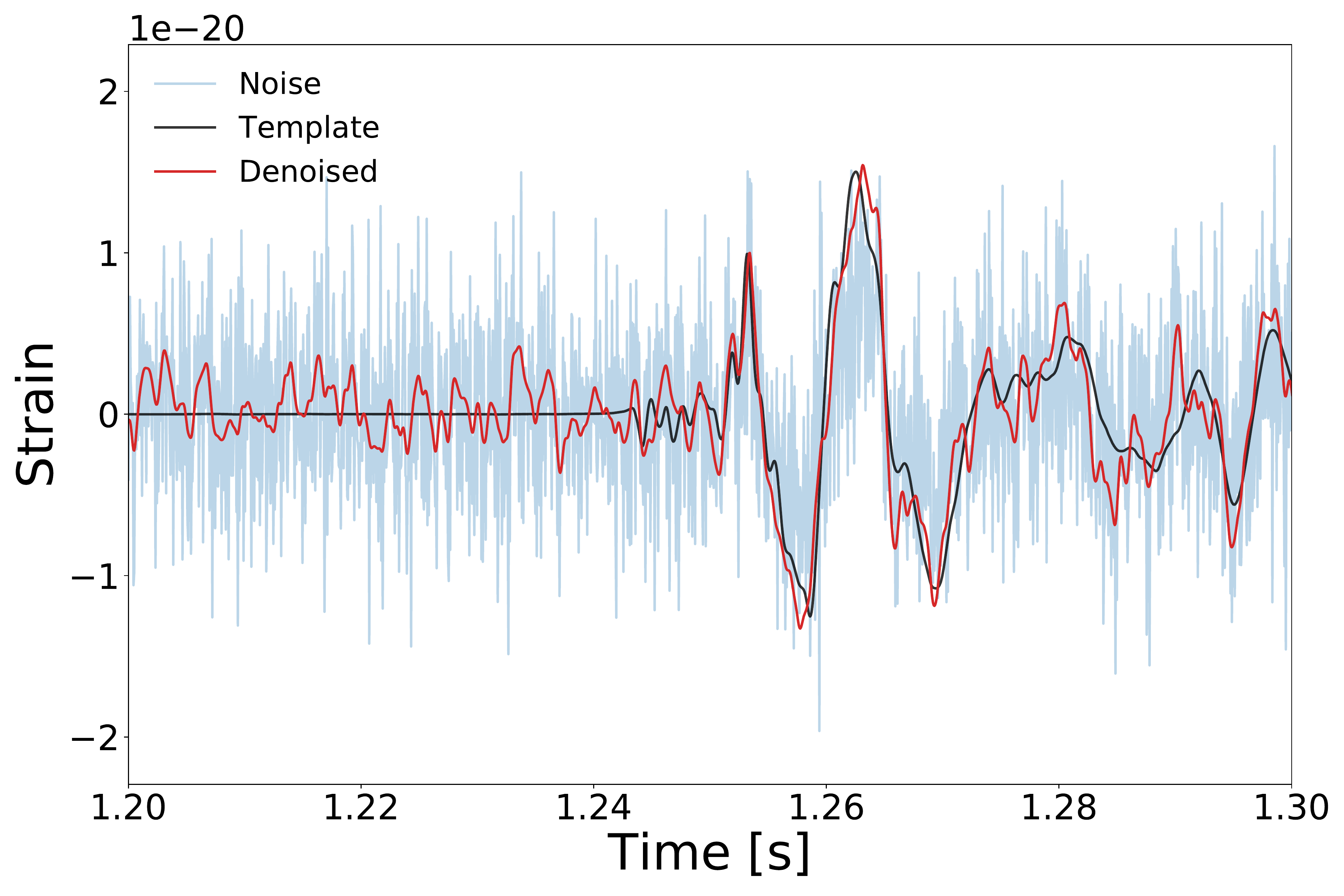}}\\
 \caption{Results of applying the rROF model to three CCSN signals from the Dimmelmeier catalog~\cite{Dimmelmeier:2008} at a distance for 10 kpc, namely 60 (upper panel), 68 (middle panel), and 96 (lower panel). Black lines indicate the original signals (after whitening) and red lines correspond to the denoised ones.}
  \label{fig:burst}
\end{figure}

We first assess our method with three signals from the CCSN catalog placed at a distance of 10 kpc and using the optimal value of $\mu$ for each case. The signals correspond to numbers 60, 68 and 98 of the Dimmelmeier catalog~\cite{Dimmelmeier:2008}. With the source at 10 kpc the signal is visible over the noise. However, despite its simplicity this is the first test the method shall pass. The results are displayed in Fig.~\ref{fig:burst}. As the distance is fixed, the strain (amplitude) of the signal depends on the particular simulation, i.e., the SNR is different for each case. Fig.~\ref{fig:burst} shows that signal 60 has the largest amplitude (top panel).

All denoised signals are very similar to the whitened templates. Signal 60 is the strongest and the denoised signal fits the template almost perfectly. The other two signals are weaker at 10 kpc and the denoising procedure leads to more oscillatory signals. The amplitude and phase of the main positive and negative peaks and the first secondary peak in all signals are well recovered. In contrast, the low amplitude damped oscillations that follow the burst (associated with the oscillation of the proto-neutron star) are lost since, due to their low amplitude, they are more affected by noise. This is inherent to the rROF model, as it preserves large gradients and disfavors small ones. 

The denoised signal shown in the middle panel of Fig.~\ref{fig:burst} is more oscillatory than the other two. In this case, a higher value of $\mu$ is required to recover its peaks properly which leads to the presence of more noise than in the other cases. This fact is something to take into account in a realistic case where the real signal is unknown, because these small oscillations could be disregarded in favor of a more regular signal and more noise removal. However, it is always possible to use a larger value of $\mu$ to recover these parts of the signal.

\begin{table}
\label{tab:results}
\caption{Values of the SSIM index for CCSN waveforms when using the optimal value of the regularization parameter for each signal, $\mu_{\rm{opt}}$, the mean value for all signals, $\bar{\mu}$, and multiple values, $\mu_{\rm{m}}$. The final column, `ref', indicates the SSIM index computed for the signal obtained after the whitening and the corresponding template.}
\begin{center}
\begin{tabular}{cccccc}
Signal & Distance &  \multicolumn{4}{c}{SSIM index} \\
& (Mpc) & [$\mu_{\rm{opt}}$] & [$\bar{\mu}$] & [$\mu_{\rm{m}}$] & [ref] \\[1ex]

 \hline
   & 5  & 0.89 & 0.83 & 0.84 & 0.39 \\
60 & 10 & 0.74 & 0.68 & 0.69 & 0.14\\
   & 20 & 0.54 & 0.44 & 0.43  &0.03 \\
\hline
   & 5  & 0.71 & 0.61 & 0.65& 0.21 \\
68 & 10 & 0.51 & 0.33 & 0.38 &0.06  \\
   & 20 & 0.31 & 0.08 & 0.11& 0.003  \\ 
\hline
   & 5  & 0.64 & 0.60 & 0.69 &0.06  \\
96 & 10 & 0.40 & 0.46 & 0.51  &0.012 \\
   & 20 & 0.23 & 0.24 & 0.29 & 0.002 \\
 \hline
\end{tabular}
\end{center}
\label{tab:burst}
\end{table}
To complete the analysis, we also compute the denoising using the mean value of the regularization parameter and the multiple regularization for all cases, where the distance is different (5, 10, and 20 kpc). The resulting values of the SSIM index are shown in Table \ref{tab:burst}. In particular, the last column of this table shows the SSIM index computed for the signal obtained after the whitening and the corresponding numerical template. Therefore, it provides a measure of the improvement obtained with the rROF method. The values of the SSIM index are computed in a 256 window centered at the position of the negative peak of the signal. These values are computed before applying the TV method in order to illustrate its performance. As expected, the values of the SSIM index become worse as the distance increases, irrespective of the type of regularization parameter employed. The comparison shows that the optimal value of the regularization parameter produces the best results in all cases. The denoising worsens when using the mean value of $\mu$, however the results are similar to those obtained with the optimal one. It thus seems possible to use the mean value for all signals and still obtain good results. Likewise, the use of multiple regularization values seems to be a good alternative too because the values of the SSIM index are very similar to the other cases. For the case of $\mu_{\rm{m}}$, the values of the SSIM index depend on the sampling of the Gaussian distribution. We have repeated the sampling several times finding similar results. It is expected that as we increase the number of samples obtained from the distribution of $\mu$ the value of the SSIM index will converge to that obtained with $\bar{\mu}$. However, multiple regularization has the advantage that the results can be analyzed separately. 

The results of Table \ref{tab:burst} also show that there is not a very strong dependence with $\mu$, i.e.~if the chosen value is of the same order of magnitude than $\mu_{\rm opt}$, the results are quite similar. This is a different behaviour with respect to the results we found in~\cite{Torres:2014} where this dependence was more critical. The reason is the use of the iterative procedure which allows to choose larger initial values of $\mu$. Slightly different values of $\mu$ will require different number of iterations to reach convergence, but the result will be similar.

To further test the performance for signals at 20 kpc, we make a complementary test. We compute the spectrogram of each signal and integrate the power for each temporal channel. Then we determine the time at which the 
maximum power is achieved. We make this calculation for the noisy and the denoised signals and compare the results with the waveform template. In all cases considered the time given by the denoised signal matches the one given by the template even if the noisy signal (after whitening) does not. 

\subsection{Binary black hole signals}
\label{BBH}

We turn next to perform the same analysis to BBH waveform signals. These signals are significantly longer than those from CCSN and are composed by three parts, inspiral, merger, and ringdown. During the inspiral, the signal amplitude and frequency increase up until merger. For our tests, as we did in~\cite{Torres:2014}, we employ signal BBH0001 from the catalog of~\cite{Mroue:2013}.
\begin{figure}[t]
{\includegraphics[width=0.45\textwidth]{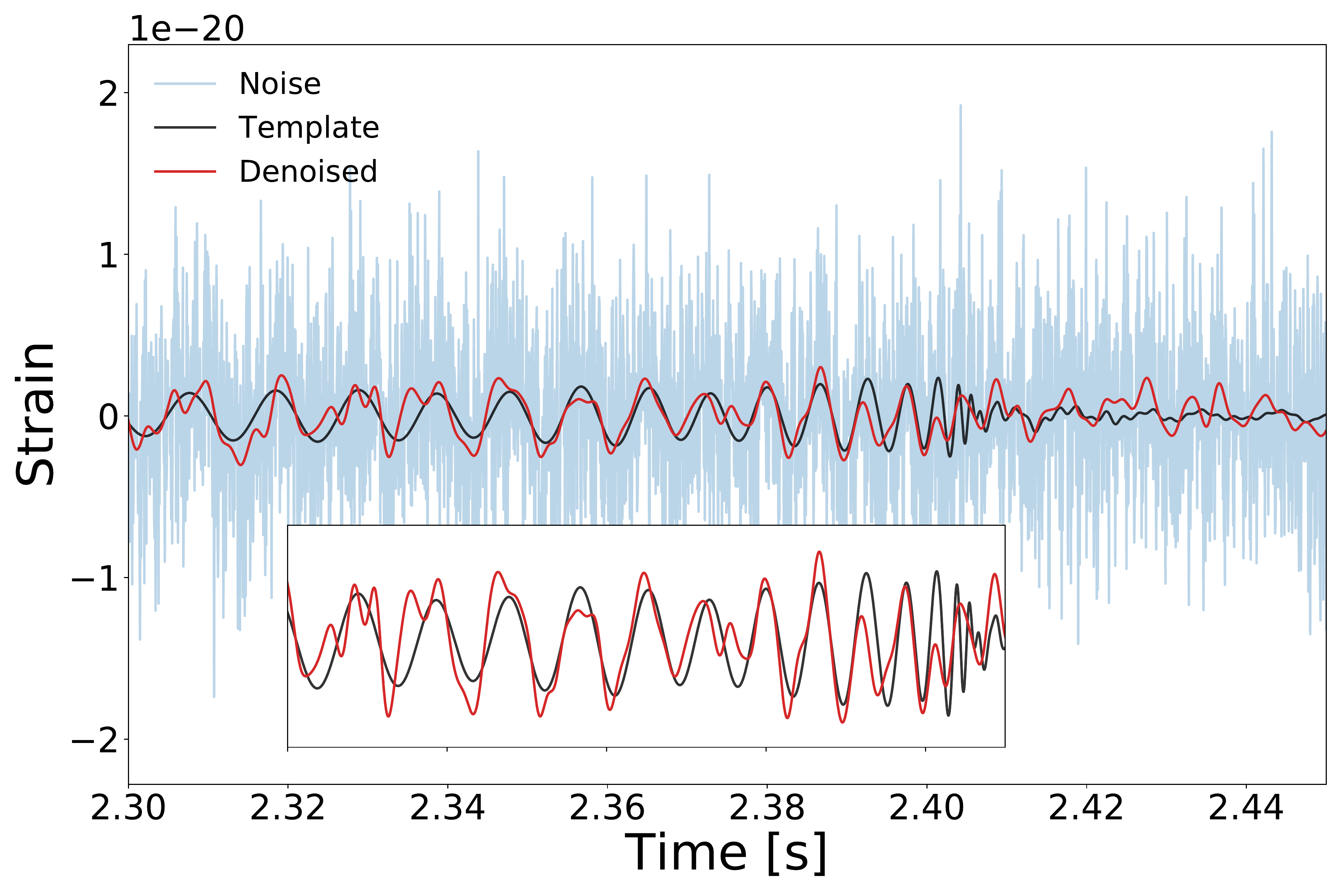}}
\caption{Denoising of signal BBH0001 of the Mrou\'e catalog~\cite{Mroue:2013} at a distance of 400 Mpc. 
The black line is the original template (after whitening) and the red line corresponds to the denoised signal.
In the inset the noise is not plotted and the area of interest is enlarged to facilitate the comparison.}
 \label{fig:BBH}
\end{figure}

The results of denoising this BBH signal placed at a distance of 400 Mpc are shown in Fig.~\ref{fig:BBH}. The last few cycles before merger (at $t\sim 2.4$ s) is the part of the signal the TV algorithm recovers best, as expected, since the value of $\mu$ is adapted to this region. Previous cycles of the inspiral are less smooth because, in general, the merger requires a higher value of $\mu$, a choice that is not optimal for the rest of the signal. 

The dependence of our results with other distances and other possible choices of the regularization parameter are reported in Table~\ref{tab:bbh}. Both $\bar{\mu}$ and $\mu_{\rm{m}}$ produce worse results than the optimal regularization parameter as the distance increases. This may happen if the optimal value for a given distance lies at the tail of the distribution shown in the histogram, as for example in the case of 800 Mpc, where $\mu_{\rm{opt}}=-2.45$ which is very different from $\bar{\mu}$. Also note that for a distance of 1 Gpc, all three choices of $\mu$ lead to similar fairly low values of the SSIM index.
\begin{table}
\label{tab:results}
\caption{Values of the SSIM index for a BBH waveform when using the optimal value of the regularization parameter for each signal, $\mu_{\rm{opt}}$, the mean value for all signals, $\bar{\mu}$, multiple values, $\mu_{\rm{m}}$, and the reference value computed for the signal obtained after the whitening and the corresponding template.}
\begin{center}
\begin{tabular}{cccccc}
Signal & Distance &  \multicolumn{4}{c}{SSIM index} \\
& (Mpc) & [$\mu_{\rm{opt}}$] & [$\bar{\mu}$] & [$\mu_{\rm{m}}$] & [ref] \\[1ex]
 \hline
& 400 & 0.43 & 0.25 & 0.25  & $4\times10^{-4}$ \\
0001 & 800 & 0.26 & 0.14 & 0.10 & $1\times10^{-4}$\\
 & 1000 & 0.11 & 0.10 & 0.10&$ 4\times10^{-5}$ \\
 \hline
\end{tabular}
\end{center}
\label{tab:bbh}
\end{table}

As the waveform produced by a BBH coalescence has significant length and variations in frequency and amplitude, the optimal value of $\mu$ is different for different parts of the signal. In Fig.~\ref{fig:BBH} and Table~\ref{tab:bbh} we have selected the values of $\mu$ that best fit the last cycles and the merger, choosing this part as the temporal window where the SSIM index is computed. However, there is no guarantee that this value will produce the best results in other parts of the signal. 

\begin{figure}[t]
	{\includegraphics[width=0.45\textwidth]{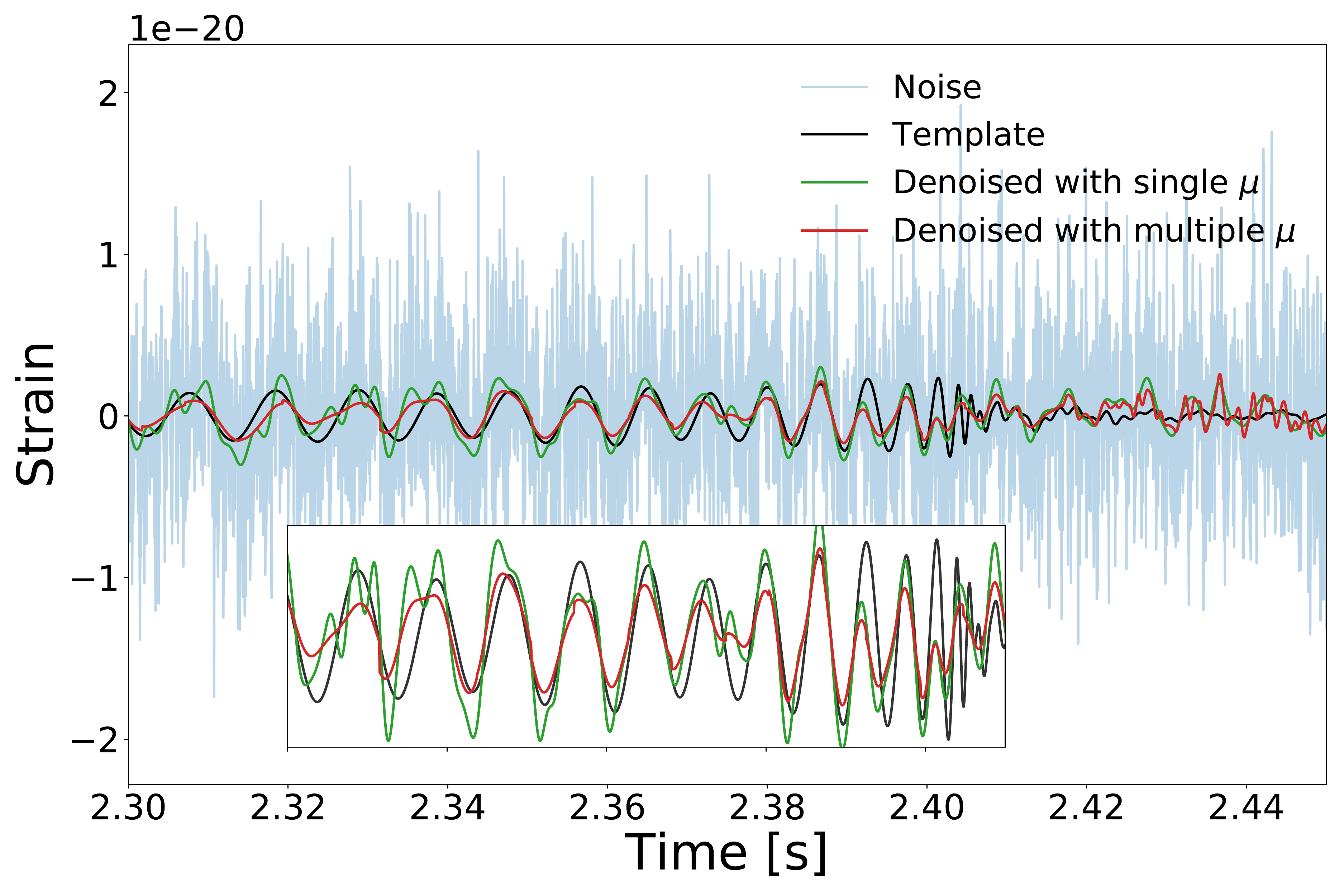}}
	 \caption{Same as Fig.~\ref{fig:BBH} but including the curve corresponding to using multiple regularization parameters.}
  \label{fig:BBH_multi}
\end{figure}

To determine the optimal value of $\mu$ in different parts of the signal we split the waveform into pieces of 256 samples and search for $\mu_{\rm opt}$ for each of the resulting windows. The denoised signal for a BBH merger at 400 Mpc is shown in Fig.~\ref{fig:BBH_multi}, both with a single value of $\mu_{\rm opt}$ and with multiple values. If we compare the results for the SSIM index restricting the comparison to the last cycles of the inspiral and the merger, the value is similar to that obtained with a single $\mu_{\rm opt}$ procedure. However, when considering the whole signal, the global value of the SSIM index significantly improves, increasing from 0.12 to 0.40.  The comparison of Fig.~\ref{fig:BBH} and Fig.~\ref{fig:BBH_multi} shows that the merger part is similar in both cases but the inspiral part is recovered better in the latter, where the value of $\mu$ has been chosen to fit each part separately. 
%

\subsection{Automatic regularization with an Artificial Neural Network}

From the results of the previous section we can devise an automatic way to find the optimal value of $\mu$ depending solely on the data at the input. The goal of this section is to present a simple way to obtain a value of $\mu$ closer to the optimum one in a realistic case, when the latter cannot be computed. This is a typical problem for machine-learning methods, in which the result is determined by the data. Machine learning algorithms have been applied to a wide (and growing) range of fields (see~\cite{Goodfellow:2016} and references there in) and a large variety of approaches are available. For our case, we implement a non-linear regressor which maps for each input window of the signal the optimal value of $\mu$ required to achieve the best denoising results. A more comprehensive analysis with different configurations of neural networks will be presented elsewhere.

\begin{figure}
	{\includegraphics[width=0.4\textwidth]{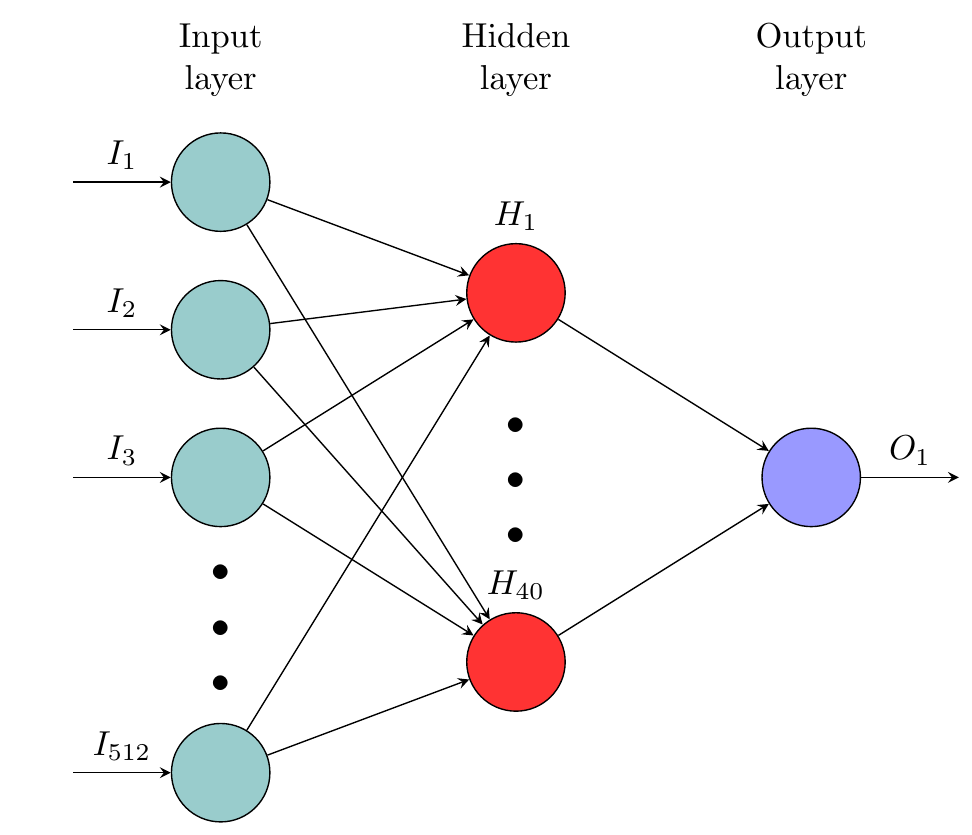}}
	 \caption{Structure of the Artificial Neural Network used to perform the regression.}
  \label{fig:net}
\end{figure}

We set up a very simple configuration with 40 neurons in one layer. The detailed structure of the network is shown in Fig.~\ref{fig:net}. Each neuron performs a linear calculation,
\begin{equation}
\label{eq:linear_calculation}
z_i = w_i^l X_i+ b_i^l ,
\end{equation}
where $w_i^l$ are the weights, $b_i^l$ is the bias parameter, $X_i$ is the input data, and $z_i$ is the output of neuron $i$ at layer $l$. Non-linearity is achieved using the so-called activation function (see~\cite{Goodfellow:2016} for details). In this case we use the well-known Relu activation, which is given by
\begin{equation}
\label{eq:relu}
f(x) = \max(0,x).
\end{equation}
The best values of the weight matrix of each layer is achieved during the training step, where for each input example the network computes the output and compares the result with the input value $\mu_{\rm{opt}}$ using the Mean Squared Error (MSE) as error quantifier. The network changes the weights iteratively to reduce the MSE. To perform this optimization procedure we use the Adam optimizer~\cite{Kingma:2014}. 

\begin{figure}
	{\includegraphics[width=0.45\textwidth]{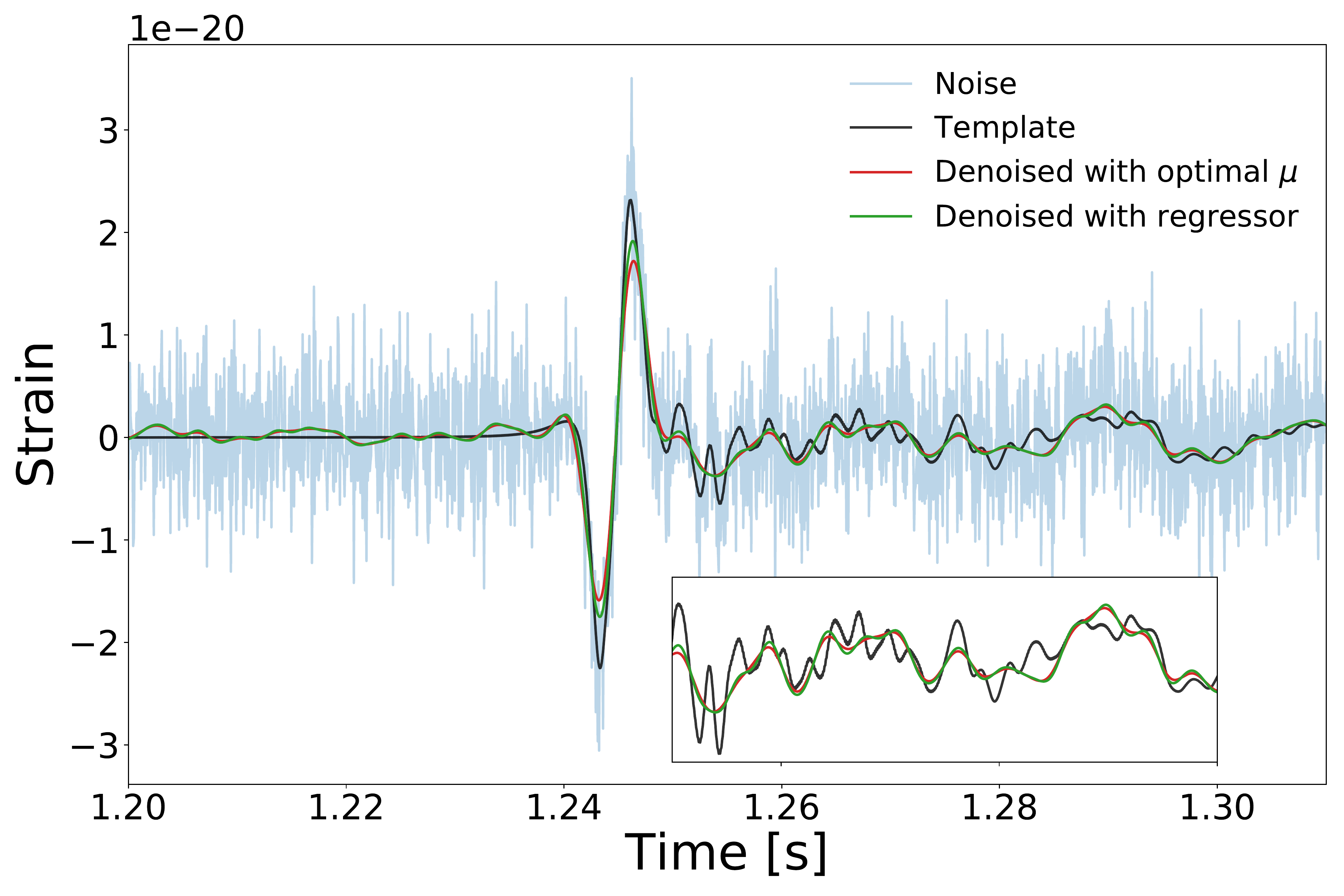}}
	{\includegraphics[width=0.45\textwidth]{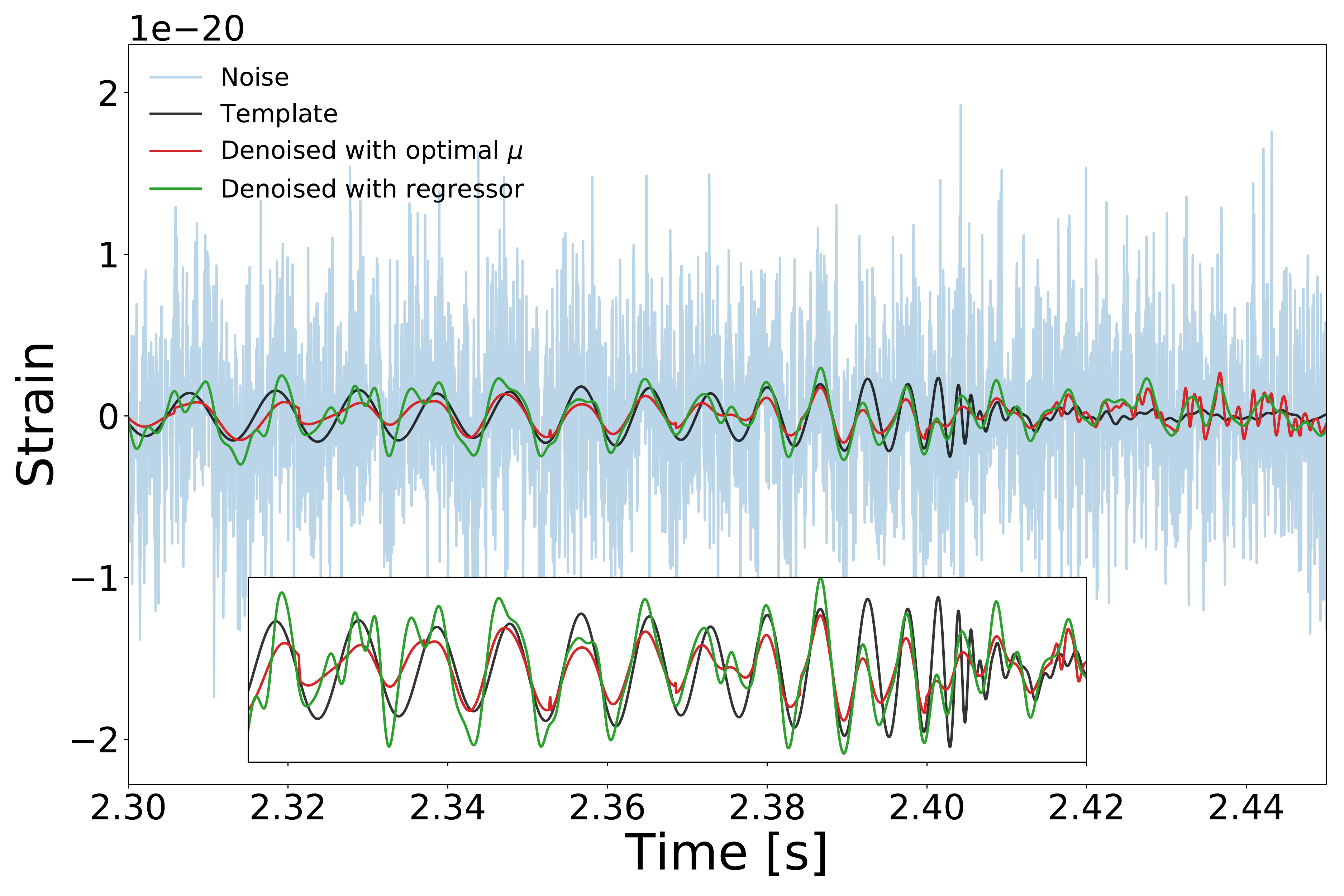}}
	 \caption{Top panel: Denoising of CCSN signal 60 at a distance of 10 kpc, showing the comparison between $\mu_{\rm opt}$ and $\mu_{\rm{reg}}$. Bottom panel: same comparison but for the BBH signal.}
  \label{fig:regression_burst}
\end{figure}

We first consider the CCSN catalog employed in Section~\ref{section:estimation}. For the training we use $10^4$ random examples of the set of 30 CCSN signals. The length of each signal example is 512 samples. The top panel of Fig~\ref{fig:regression_burst} displays the results of applying the regularization parameter determined by the regressor to the CCSN signal 60 of~\cite{Dimmelmeier:2008} at 10 kpc. As the figure shows, the results achieved with the optimal value of $\mu$ and with the one given by the regressor are very similar (both curves almost completely overlap). In addition, Table \ref{tab:burst_regression} reports the values of the SSIM index after applying the regression procedure to three CCSN signals at three different distances, and the values for $\mu_{\rm opt}$ for reference. The results are different to the ones shown in Table \ref{tab:burst} because of the different window used. Here we 
use a sliding window to denoise the entire signal and thus the windows are not exactly the same. For most cases the values of the SSIM index given by the regressor are similar to the optimals. 

\begin{table}
\label{tab:results}
\caption{Comparative results between using the optimal value of $\mu$ and the value given by the regressor $\mu_{\rm{reg}}$, for three CCSN signals.}
\begin{center}
\begin{tabular}{cccc}
Signal & Distance &  \multicolumn{2}{c}{SSIM index } \\
& (kpc) & $[\mu_{\rm{opt}}]$ & $[\mu_{\rm{reg}}]$ \\[1ex]
 \hline
  & 5 & 0.78 & 0.73 \\
60 & 10 & 0.64 & 0.63 \\
  & 20 & 0.46 & 0.38 \\
\hline
  & 5 & 0.25 & 0.26 \\
68 & 10 & 0.16 & 0.16  \\
 & 20 & 0.10 & 0.05  \\
\hline
  & 5 & 0.62 & 0.51 \\
96 & 10 & 0.55 & 0.51 \\
  & 20 & 0.46 & 0.46 \\
 \hline
\end{tabular}
\end{center}
\label{tab:burst_regression}
\end{table}

Finally, we train the regressor with signals from the BBH catalog and we repeat the analysis done in Section \ref{BBH}. Specifically, we train the network with $10^4$ random examples of the set of 30 BBH signals, at 5 different GPS times and 3 distances, and a window of 512 samples. The value of the SSIM index is 0.41 for the optimal regularization parameter, 0.36 for the regressor in each window, and 0.2 if we use for all the signal $\mu_{\rm opt}$ computed at merger. The comparison is shown in the bottom panel of Fig~\ref{fig:regression_burst}.



\section{Summary}
\label{section:summary}

This paper has extended the work we initiated in~\cite{Torres:2014} to denoise gravitational-wave signals using total-variation methods. We have assessed these techniques in real noise conditions, injecting numerical-relativity waveforms from CCSN and BBH mergers in data from the first observing run of Advanced LIGO. We have shown that TV methods remove noise irrespective of the type of signal. The denoising procedure is performed in two steps. First, we apply a whitening procedure to remove lines and to flaten the noise spectrum, and next we apply the TV method. The quality of the results depends on the value of the regularization parameter $\mu$ of the ROF model. Therefore, we have to perform an heuristic search for the Lagrange multiplier that produces the best results. To improve the statistics, we have computed the optimal value of $\mu$ for 30 different signals from a CCSN catalog~\cite{Dimmelmeier:2008} and for another 30 from a BBH merger catalog~\cite{Mroue:2013}, placing the signals at various distances. The histograms have shown that the interval of optimal values of $\mu$ is not very wide. However, the use of an iterative procedure reduces somewhat the denosing dependence on $\mu$ and has allowed us to obtain similar results with a larger span of values of $\mu$ than in our first paper~\cite{Torres:2014}. 

In a realistic situation, however, the original signal is not known and it is not possible to determine the optimal value of the regularizer. Therefore, in this work we have expanded the analysis by testing additional ways to perform the denoising using a general value of $\mu$. The first ideas are based on the results of the histograms. We have shown that using the mean value can suffice in most cases. Another approach uses 20 different values of $\mu$ to compute the mean, yielding similar results. Multiple-$\mu$ denoising allows to compare the results at different scales, to process them separately and to combine them to obtain the best results. However, even though the method is very fast (on average it takes about 0.5 s to denoise 3 s of signal with the iterative procedure), it requires multiple blind selections of $\mu$, which in some cases might not be adequate. For this reason, we have also tested the use of a neural network to determine the value of the regularization parameter. We plan to  further explore this approach in a future investigation.

For the case of long-duration signals such as BBH waveforms our results have shown that a single value of the regularization parameter does not provide a good enough denoising across the entire signal. Instead, combining the results using optimal values of $\mu$ adapted to different parts of the signal improves the results. We have generalized this procedure by employing a regressor implemented with an artificial neural network of 40 neurons in one layer. We have shown that this machine-learning approach leads to results similar to those obtained with the optimal regularization parameter. Therefore, it is worth to combine TV methods with machine learning techniques to improve the results and obtain the Lagrange multiplier $\mu$ through an automatic search. 

This paper provides further evidence that TV methods can be useful in the field of Gravitational-Wave Astronomy as a tool to remove noise. They can be used in a preprocessing step before applying other common techniques of gravitational-wave data analysis. However, on their own they cannot constitute a standalone pipeline since, as they do not use any information about sources, they cannot detect, classify or extract physical information. We plan to investigate a combined strategy and test if the application of TV-denoising can improve the results of other approaches, for example by reducing the uncertainties in Bayesian methods or reducing the false alarm rate. Furthermore, we want to keep exploring the combination of TV methods with machine learning techniques. More precisely, we will further explore the determination of $\mu$ via machine learning regressors and we will work on improving the denoising algorithm itself by using multilayer structures typical of Deep Learning. Our findings will be presented elsewhere.

\section*{Acknowledgements}
Work supported by the Spanish MINECO (AYA2015-66899-C2-1-P), by the Generalitat Valenciana (PROMETEOII-2014-069), and by the European Gravitational Observatory (EGO-DIR-51-2017).


\bibliographystyle{apsrev}
\bibliography{references}

\end{document}